# Feasibility Analysis and Preliminary Design of ChipSat Entry for In-situ Investigation of the Atmosphere of Venus

**Salvatore Vivenzio[a]\*, Dan Fries[b], Chris Welch[c]**

[a] *International Space University, 1 Rue Jean-Dominique Cassini, 67400 Illkirch-Graffenstaden, France,*
salvatore.vivenzio@community.isunet.edu

[b] *Initiative for Interstellar Studies, Bone Mill, New Street, Charfield, GL12 *ES, United Kingdom*
dfries@gatech.edu

[c] *International Space University, 1 Rue Jean-Dominique Cassini, 67400 Illkirch-Graffenstaden, France,*
chris.welch@isunet.edu

\* Corresponding Author

## Abstract

Recent miniaturization of electronics in very small, low-cost and low-power configurations suitable for use in spacecraft have inspired innovative small-scale satellite concepts, such as ChipSats, centimeter-scale satellites with a mass of a few grams. These extremely small spacecraft have the potential to usher in a new age of space science accessibility. Due to their low ballistic coefficient, ChipSats can potentially be used in a swarm constellation for extended surveys of planetary atmospheres, providing large amounts of data with high reliability and redundancy. We present a preliminary feasibility analysis of a ChipSat planetary atmospheric entry mission with the purpose of searching for traces of microscopic lifeforms in the atmosphere of Venus. Indeed, the lower cloud layer of the Venusian atmosphere could be a good target for searching for microbial lifeforms, due to the favourable atmospheric conditions and the presence of micron-sized sulfuric acid aerosols. A numerical model simulating the planetary entry of a spacecraft of specified geometry, applicable to any atmosphere for which sufficient atmospheric data are available, is implemented and verified. The results are used to create a high-level design of a ChipSat mission cruising in the Venusian atmosphere at altitudes favorable for the existence of life. The paper discusses the ChipSat mission concept and considerations about the spacecraft preliminary design at system level, including the selection of a potential payload.

## Nomenclature

| Symbol | Meaning |
|---|---|
| $A_c$ | Cross Sectional Area |
| $a_s$ | Speed of Sound |
| $BC$ | Ballistic Coefficient |
| $C_D$ | Drag Coefficient |
| $C_L$ | Lift Coefficient |
| $D$ | Drag Force |
| $g$ | Gravity Acceleration |
| $h$ | Altitude |
| $Kn$ | Knudsen Number |
| $L$ | Lift Force / Characteristic Length |
| $M$ | Mass |
| $Ma$ | Mach Number |
| $R$ | Position Vector / Radius |
| $T$ | Atmospheric Temperature |
| $t$ | Time |
| $v$ | Velocity |
| $v_w$ | Normal Component of Thermal Velocity |
| $W$ | Weight |
| $\alpha$ | Angle of Attack |
| $\Gamma$ | Local Specific Gas Constant |
| $\gamma$ | Flight Path Angle |
| $\varepsilon$ | Emissivity |
| $\eta_n$ | Molecular Accommodation Coefficient along the Normal Direction |
| $\eta_t$ | Molecular Accommodation Coefficient along the Tangential Direction |
| $\lambda$ | Mean Free Molecular Path |
| $\rho$ | Atmospheric Density |

**Acronyms/Abbreviations**

AEM    Atmospheric Entry Model
ARD    Atmospheric Reentry Demonstrator
COTS    Commercial-Off-The-Shelf
ISM    Industrial, Scientific and Medical (radio bands)
IR    Infrared
L-M    Liquid Chromatography-Mass Spectrometry
LOC    Lab-on-a-Chip
MEMS    Micro-Electro-Mechanical Systems
PCB    Printed Circuit Board
SMD    Surface Mounted Device
USSA    U.S. Standard Atmosphere
UV    Ultraviolet

## 1. Introduction

This paper proposes the use of the ChipSat technology for extended surveys of planetary atmospheres. ChipSats are centimeter-scale satellites with a mass of a few grams that, thanks to their low cost and low time for development (from years to months), reduce the entry barriers for spacecraft development, making space exploration affordable for education institutions and for a large range of space amateurs. The current state of the art for ChipSats is well represented by the "Sprites" spacecraft, the smallest space probes built on a single circuit board. More than a hundred Sprite 3.5-by-3.5 centimeter ChipSats, developed by Zack Manchester (Stanford University) in collaboration with the NASA Ames Research Center [1], have been deployed on March 19, 2019 from the Kicksat-2 CubeSat carrier. Radio signals from this formation have been received to confirm this release [2]. A similar spacecraft is the Monarch ChipSat developed by Cornell University that, according to Adams and Peck [3], has advanced beyond the Sprites effort, but it has not been launched yet into space.

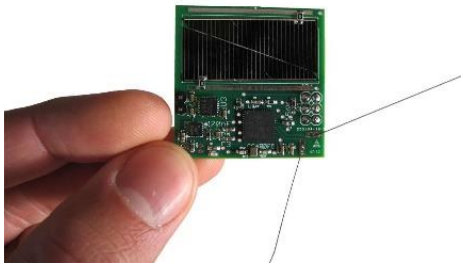

*Figure 1. Sprites ChipSat sample* [4]

The use of these small-sized satellites opens the doors to new space mission concepts. Some have already been proposed in the past, such as swarms of ChipSats equipped with a wide range of MEMS and sensors, performing planetary atmospheres characterization studies [5], or asteroid composition analyses [6], producing simultaneously a large amount of data over a large volume of space, assuring high spatial resolution and redundancy. A mission based on a large number of ChipSats provides a high degree of robustness and high probability of success, without using any redundant subsystems and high-reliability components, unlike traditional spacecraft [7], [8].

Therefore, ChipSats are proposed for a mission to perform in-situ investigation of Venus' atmosphere. The interest in such a challenging missions derives from the results of several astrobiology studies suggesting the potential presence of microbial life in the Venusian atmosphere.

It is well known that the atmospheric conditions on Venus surface are very harsh. The temperature values are around 460 °C and the atmospheric pressure is 92 times the Earth's surface pressure. Nevertheless, according to a recent study published in the journal *Astrobiology* in 2018 by Limaye et al. [9], the lower cloud layer of the Venusian atmosphere could be a good target for searching for microbial lifeforms (e.g. bacteria), due to the favourable atmospheric conditions and the presence of micron-sized sulfuric acid aerosols. In the past, it is likely that Venus hosted water for a relatively short period, between 650 million to 2 billion years, before it boiled off into the clouds. Nonetheless, this could still be a period sufficiently long to develop microbial life, considering that Earth needed less than a billion years, according to Kluger [10]. As terrestrial bacteria have routinely been sampled in the upper atmosphere, surviving at those atmospheric conditions, the potential Venusian bacteria could have done the same [10]. It is evident that atmospheric conditions, radiation levels and the presence of available water are the major limitations for habitability. However, this is not the case for the bottom cloud layer of Venus, which presents rather favourable conditions, with a temperature of ~60 °C and a pressure of ~1 atm, and a UV radiation flux comparable with that of the Archean Earth's surface, when photosynthetic life was considered to be present on our planet. The major limitation is given by the availability of liquid water in the clouds, which still has to be assessed [9]. Moreover, in 1993, Grinspoon et al. [11] proposed that the ultraviolet spatial contrasts discovered in the early part of the 20th century in Venus' cloud layer (whose causes are still unknown), could be justified by the presence of "unknown ultraviolet absorbers", which may represent one of the signs for presence of life (biomarkers). Current models include sulfur dioxide and iron chloride as the UV absorbers, which could suggest the possibility of a potential iron- and sulfur-centered metabolism in the clouds [9]. Furthermore, Limaye et al. [9] mention that Venus' lower clouds present mass loading estimates comparable to the upper biomass value for the primary biological aerosols in Earth's atmosphere, which are dominated by bacteria. In the same work the authors suggest that the estimated sizes of Venus' clouds particles are comparable to those of the Earth's biological aerosols particles (from nanometer to submillimeter), opening the possibility that the clouds may similarly harbor microbial life.

This paper assesses the feasibility of a mission to investigate the Venusian atmosphere using ChipSats, by developing an *Atmospheric Entry Model* (AEM) able to compute and simulate the planetary entry conditions of a spacecraft of specified geometry in the atmosphere of any planet for which sufficient atmospheric data are available.

After validating the model, the results of the simulation of a ChipSat re-entry in Venus' atmosphere are used to estimate the *useful time* such a small spacecraft could have to perform its scientific analysis at a selected range of altitudes, where recent astrobiology studies assess it is more likely to find microbial life.

Section 2 presents an overview of the developed AEM, describing the models used for the computation of the planetary atmospheric parameters and the aerodynamic coefficients, as well as the mathematical models implemented for the spacecraft translational dynamics.

Section 3 deals with the verification of the results obtained from the AEM for an Earth atmospheric entry, considering both a traditional-sized spacecraft and a ChipSat spacecraft configuration.

Section 4 presents and discusses the results concerning a Venus atmospheric entry of a ChipSat in face-on configuration, obtained using the AEM. Furthermore, a parametric analysis is presented to maximize the useful time of a ChipSat entry in Venus' atmosphere.

In Section 5, the mission concept proposed to perform in-situ investigation of the lower cloud layer of Venus' atmosphere is described at high level, together with a discussion of the potential design of the spacecraft.

## 2. The Atmospheric Entry Model (AEM)

The *Atmospheric Entry Model* is a numerical model used to simulate the planetary entry conditions of a spacecraft of specified geometry in the atmosphere of any planet for which sufficient atmospheric data are available.

The AEM computes all the main characteristic parameters (including the altitude-dependent profiles of velocity, flight path angle, and acceleration load) of an atmospheric entry mission, according to the manual input selected by the user. In order to perform the planetary atmospheric entry simulation, the AEM uses Ordinary Differential Equations (ODEs) for the spacecraft translational dynamics, which are integrated using the MATLAB ODE solver *ode15s*. The mathematical model is described in Section 2.3.

The manual inputs required for running a simulation are listed below:

- **Planet**: according to the planet selected, the AEM reads a specific datasheet containing the atmospheric data for the planet's atmosphere. More details about the atmospheric models implemented in the AEM are presented in Section 2.1.
- **Spacecraft Configuration**: the AEM has a set of predefined spacecraft configurations, including regular Apollo-style and micro-scale spacecraft (ChipSats). The versatility of the model enables the user to add a new configuration if needed.
- **Angle of Attack** ($\alpha$): the AEM takes into account a constant attitude angle which has to be specified by the user.
- **Altitude Range**: the range of altitudes at which the AEM solves the ODEs to perform the simulation.

It is important to mention that the AEM has the following intrinsic limitations:

- It does not include an aerothermodynamics model. Nevertheless, results available in literature are used to complement the translational dynamics simulations presented in this paper.
- The rotational motion is neglected.
- A model for the lift coefficient is implemented for the free molecular flow regime only. Nevertheless, the simulations presented regard spacecraft configurations for which the lift is negligible.

### 2.1 Atmospheric Models for Earth and Venus

The *Atmospheric Entry Model* is able to use different atmospheric models, according to the planet selected by the user. Currently, it includes the U.S. Standard Atmosphere 1976 model for Earth, presented in [12], and the Venus International Reference Atmosphere (VIRA) model for Venus, described by Kliore, Moroz and Keating [13].

The U.S. Standard Atmosphere 1976 is an idealized, steady-state representation from the Earth's surface up to 1000 km, assuming a moderate solar activity. The AEM uses this model to calculate the density and temperature profiles varying with the altitude.

The Venus International Reference Atmosphere (VIRA) model was developed from data observed by Magellan and other Venus orbiters and entry probes. The model includes the density and temperature profile up to 150 km. In particular, the deep and medium atmosphere data (up to 33 km and 100 km respectively) are extracted from [14], whilst the upper atmosphere data (from 100 to 150 km) from [15].

Furthermore, in order to define the flow regime experienced by the spacecraft during entry, the main free molecular path reported respectively in U.S. Standard Atmosphere 1976 and VIRA models are used to compute the Knudsen number profile for Earth and Venus. The Knudsen number is defined as $Kn = \lambda/L$, where $\lambda$ is the mean free molecular path, and $L$ is the characteristic length of the spacecraft. Table 1 shows the limit values of $Kn$, discriminating the different flow regimes.

Table 1: Different flow regimes depending on the Knudsen number

| $Kn < 0.1$ | Continuum Flow |
| --- | --- |
| $0.1 \leq Kn \leq 10$ | Transitional Flow |
| $Kn > 10$ | Free Molecular Flow |

Figure 2 and Figure 3 show the variation of the Knudsen number with the altitude respectively for Earth's and Venus' atmospheres (reproduced with the AEM), considering the case of a ChipSat with a side length of 1 cm, which is used as the characteristic length.

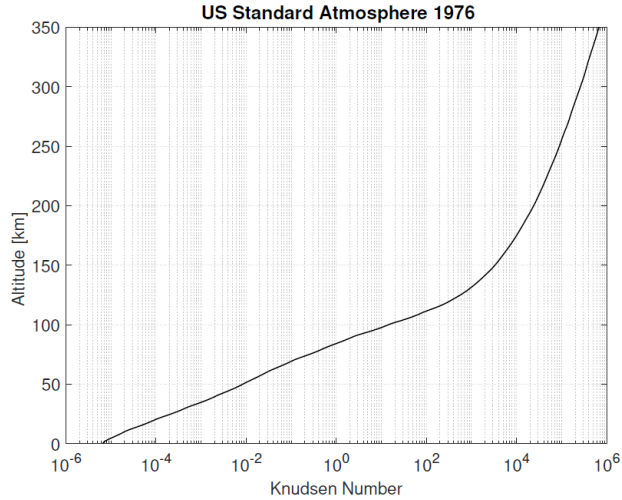

Figure 2. Knudsen number vs Altitude for a ChipSat with a 1 cm side length in Earth's Atmosphere

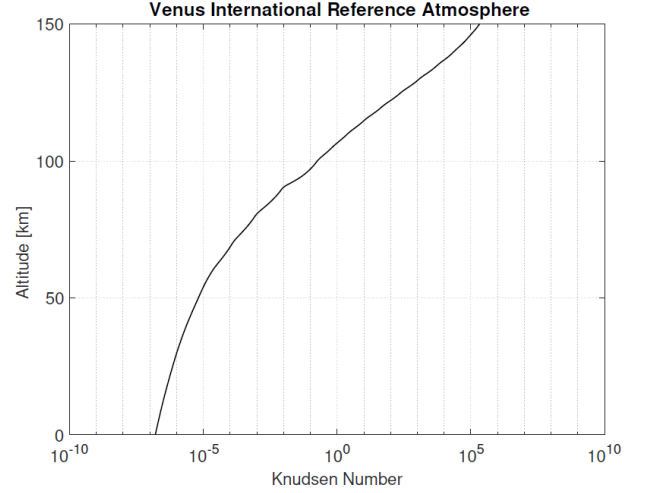

Figure 3. Knudsen number vs Altitude for a ChipSat with a 1 cm side length in Venus's Atmosphere

2.2 Aerodynamic Coefficients

The mathematical models for the drag and the lift coefficients are chosen with the main purpose of this work in mind, which is to simulate a planetary atmospheric entry for a ChipSat. In a simplified manner, this very small spacecraft can be thought of as a flat plate with side-lengths and a thickness, respectively, on the order of centimeters and milli- to micrometers.

The aerodynamic coefficients strongly depend on the flow regime the spacecraft experiences during the entry, determined by the Knudsen number, as mentioned in Section 2.1. This is a fundamental aspect to be considered when simulating the flight of such a small spacecraft.

The models for both $C_D$ and $C_L$ in free molecular flow are given by Storch [16], whose expressions ignore spinning and tumbling body effects. They are shown in equations (2.2.a) and (2.2.b).

$$C_D = 2\left[\eta_t + \eta_n \frac{v_w}{v}\sin(\alpha) + (2 - \eta_n - \eta_t)\sin^2(\alpha)\right]\sin(\alpha) \quad (2.2.a)$$

$$C_L = \left[\eta_n \frac{v_w}{v} + (2 - \eta_n - \eta_t)\sin(\alpha)\right]\sin(2\alpha) \quad (2.2.b)$$

$v_w = \sqrt{\frac{\pi \Gamma T}{2}}$ is the normal component of the thermal velocity ($T$ is the local atmospheric temperature, $\Gamma$ is the local specific gas constant), whilst $\eta_n$ and $\eta_t$ are dimensionless coefficients modelling the molecular accommodation in the plate's normal and tangential directions.

In the continuum flow regime the AEM considers a constant $C_D = 1.28$, which is an empirical result obtained for flat plates in subsonic regime, reported by [17]. The results discussed in Section 4.1 confirm the validity of the assumption of subsonic flow when the spacecraft enters the continuum regime in Venus' atmosphere.

In the transitional flow regime, the $C_D$ is computed by interpolating between the value used for continuum flow and the value calculated at an altitude corresponding to $Kn = 10$. The latter being considered the boundary value between free molecular flow and transitional flow regimes.

Figure 4 and Figure 5 show the simulated drag coefficient variation during an Earth atmospheric entry for a ChipSat in a face-on configuration ($\alpha = 90°$). In this case the free molecular flow and transitional flow boundary occurs at an altitude of about 97 km (for Earth) and 115 km (for Venus), whilst the transitional flow and continuum flow boundary occurs at about 69 km (for Earth) and 97 km (for Venus).

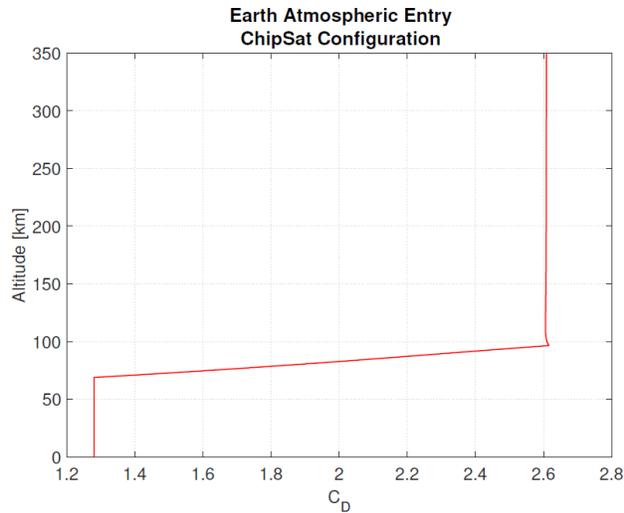

Figure 4. $C_D$ profile for a ChipSat Earth atmospheric entry

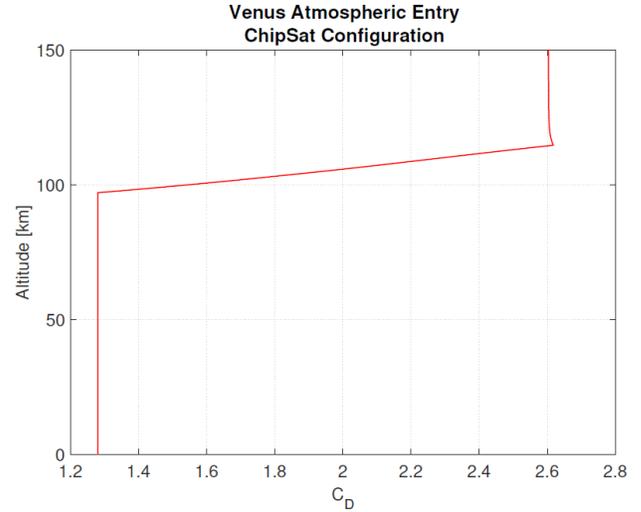

Figure 5. $C_D$ profile for a ChipSat Venus atmospheric entry

Regarding the lift coefficient, no mathematical models for the continuum and transitional flow regimes have been implemented yet, since the main purpose of the analyses presented in this work is the study of a ChipSat in face-on configuration ($\alpha = 90°$), for which the lift force effects are negligible.

## 2.3 Translational Dynamics Model

The mathematical model for the translational dynamics consists of two Ordinary Differential Equations (ODEs), which describe the spacecraft motion along the two directions tangential and normal to the curving trajectory in the point P, as represented in Figure 6, where the spacecraft is as assumed to be a mass-point.

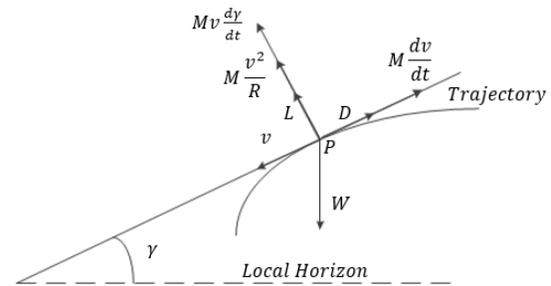

Figure 6. Scheme of the forces acting on the spacecraft

The two equations, which are reported by Monti and Zuppardi [18], are functions of two time-varying parameters:

- The spacecraft **velocity** $V(t)$.
- The **flight path angle** $\gamma(t)$ (angle between the direction tangential to the trajectory and the local horizontal direction).

The two ODEs are reported below, respectively for the tangential (2.3.a) and the normal (2.3.b) directions:

$$M\frac{dv}{dt} + D - W\sin(\gamma) = 0 \quad (2.3.a)$$

$$MV\frac{d\gamma}{dt} + L + M\frac{v^2}{R} - W\cos(\gamma) = 0 \quad (2.3.b)$$

Once defined $A_c$, $M$, $W$, and $R$ as the spacecraft cross sectional area, mass, weight and position vector, the forces acting on the body are described in Table 2.

*Table 2. Forces acting on the spacecraft*

| | |
|---|---|
| $D = \frac{1}{2}\rho v^2 C_D A_c$ | Drag force acting on the spacecraft |
| $L = \frac{1}{2}\rho v^2 C_L A_c$ | Lift force acting on the spacecraft |
| $M\frac{v^2}{R}$ | Centrifugal force due to the curvature of the trajectory |
| $Mv\frac{d\gamma}{dt}$ | Inertial force due to the variation of the path angle $\gamma$ |

Expressing the altitude $h$ as a function of time ($dh = -v\sin(\gamma)dt$), it is possible to reformulate equations (2.3.c) and (2.3.d) using $h$ as independent variable:

$$\frac{dv}{dh} = \frac{1}{2}\frac{\rho v}{\sin(\gamma)}\frac{A_c C_D}{M} - \frac{g}{v} \quad (2.3.c)$$

$$\frac{d\gamma}{dh} = \frac{1}{2}\frac{\rho}{\sin(\gamma)}\frac{A_c C_L}{M} + \frac{1}{R\sin(\gamma)} - \frac{g}{v^2\tan(\gamma)} \quad (2.3.d)$$

Defining $R_0$ and $g_0$ respectively as the planet's radius and the surface gravitational acceleration, the magnitudes of the position vector $R$ and the gravity vector $g$ can be also expressed as a function of the altitude:

$$R(h) = R_0 + h \quad (2.3.e)$$

$$g(h) = g_0 \frac{R_0^2}{(R_0 + h)^2} \quad (2.3.f)$$

Moreover, as described in Section 2.2, it is possible to express the drag coefficient $C_D$ and the lift coefficient $C_L$ as a function of $V$ and $\alpha$, using different models according to the flow regime the spacecraft experiences during the entry.

According to the planet selected by the user, the AEM reads the atmospheric datasheet coming from one of the models described in Section 2.1, to obtain the density and the temperature profiles $\rho = \rho(h)$ and $T = T(h)$, and the Knudsen number profile $Kn = Kn(h)$ (to determine the actual flow regime). Then, assuming a constant angle of attack $\alpha$ defined by the user, the AEM is able to solve the ODEs and obtain the spacecraft's entry trajectory parameters.

## 3. Numerical Model Verification

The translational dynamics model of the *Atmospheric Entry Model* presented in Section 2.3 is verified considering the case of an Earth atmospheric entry performed with two different spacecraft configurations:
- A traditional-sized spacecraft, the Atmospheric Reentry Demonstrator (ARD) vehicle. The data used as a benchmark for the verification are reported by Monti and Zuppardi [18].
- A generic ChipSat spacecraft. In this case, the data used as a benchmark come from Atchison, Manchester and Peck [5].

In both cases, the model used for the Earth's atmosphere is the U.S. Standard Atmosphere 1976, presented in Section 2.1.

### 3.1 Verification for a Traditional-sized Spacecraft Earth Atmospheric Entry

The Atmospheric Reentry Demonstrator (ARD) vehicle is characterized by a 2,800 kg mass and a main section with a 2.8 m diameter. The initial values used for the simulation are the same as those reported by [18]. They are shown in Table 3:

*Table 3. Initial values for the simulation* [18]

| | |
|---|---|
| Initial Altitude | **120 $km$** |
| Initial Velocity | 7.5 $km/s$ |
| Initial Path Angle | 1.4° |

The verification presented in this section considers constant values for the drag coefficient and the lift coefficient, to match the assumptions made by [18]. They are shown in Table 4:

*Table 4. Aerodynamic Coefficients for the simulation* [18]

| | |
|---|---|
| Lift Coefficient | **$C_L = 0$** |
| Drag Coefficient | $C_D = 1$ |

Figure 7 shows the verification results for the ARD vehicle, in terms of altitude time history, velocity time history:
- The black-solid curve represents the benchmark for the verification. The data are reported in [18].
- The yellow-dashed curve represents the results obtained using the AEM for the face-on configuration ($\alpha = 90°$), using the assumptions in Table 4.

It is evident from the plots reported below that for each parameter evaluated, the AEM results match perfectly those of [18], confirming the verification of the presented model within the given constraints.

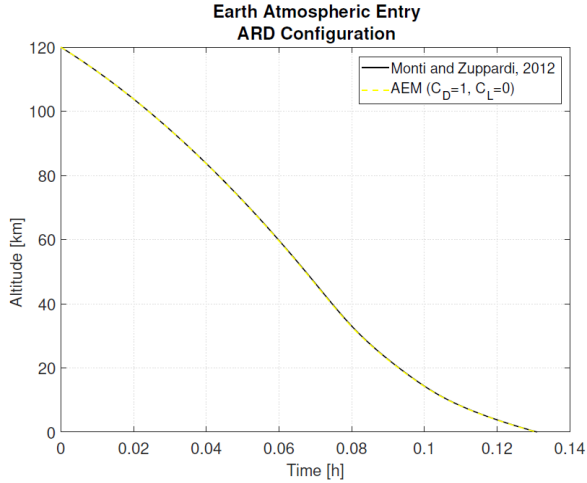

*Figure 7. Altitude time history verification for the ARD vehicle*

### 3.2 Verification for a ChipSat Spacecraft Earth Atmospheric Entry

The test case for this verification is an Earth atmospheric entry mission performed by a generic ChipSat spacecraft, thought as a silicon flat plate, whose structural specifications are listed in Table 5.

*Table 5. ChipSat Specifications*

| Side Length | **1 $cm$** |
|---|---|
| Thickness | 25 $\mu m$ |
| Material Density | 2330 $kg/m^3$ |

The initial values used for the simulation are the ones reported by [5] and are listed in Table 6.

*Table 6. Initial values for the simulation*

| Initial Altitude | **350 $km$** |
|---|---|
| Initial Velocity | 7.697 $km/s$ |
| Initial Path Angle | 0° |

The case selected concerns a ChipSat in a face-on configuration ($\alpha = 90°$), thus the lift is negligible, whilst $C_D$ assumes the profile represented in Figure 4 in Section 2.2, varying according to the flow regime experienced by the ChipSat during entry.

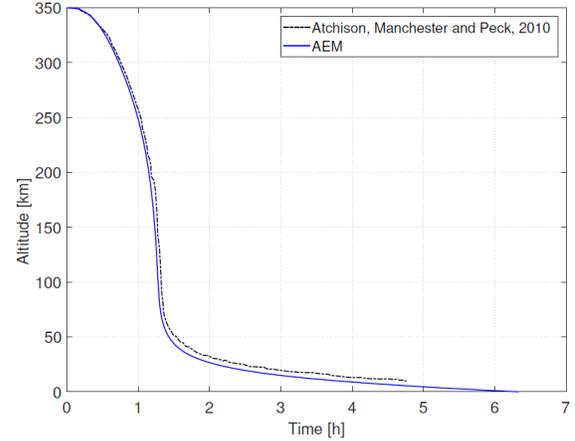

*Figure 8. Altitude time history for a face-on entry of a ChipSat*

Figure 8 shows the comparison between the altitude time history computed by the AEM model and the benchmark results coming from [5].

The AEM's trajectory implies that the ChipSat would dive into the atmosphere slightly faster than the benchmark case. These small differences in the results were expected since the drag coefficient model used in the AEM are slightly different from those reported by [5], which uses the expression of $C_D$ for the free molecular flow regime at every altitude. Nevertheless, it is evident that the trends of the two curves are almost the same, and they match almost perfectly at higher altitudes (in free molecular flow regime). This implies a positive outcome of a simple qualitative verification.

### 4. Results and Feasibility Analysis of a ChipSat Entry in the Venusian Atmosphere

#### 4.1 Results of a Venus Atmospheric Entry Performed with a ChipSat

The objective is to demonstrate the ability of the developed tool to compute the fundamental parameters of the spacecraft trajectory for a Venus atmospheric entry and to provide results supporting the parametric analysis performed in Section 4.2.

The specifications of the ChipSat selected for the Venus atmospheric entry mission are listed in Section 3.2, Table 5. The initial values used for the simulation are reported in Table 7.

*Table 7. Initial values for the simulation*

| Initial Altitude | 150 $km$ |
|---|---|
| Initial Velocity | 7.237 $km/s$ |
| Initial Path Angle | 0° |

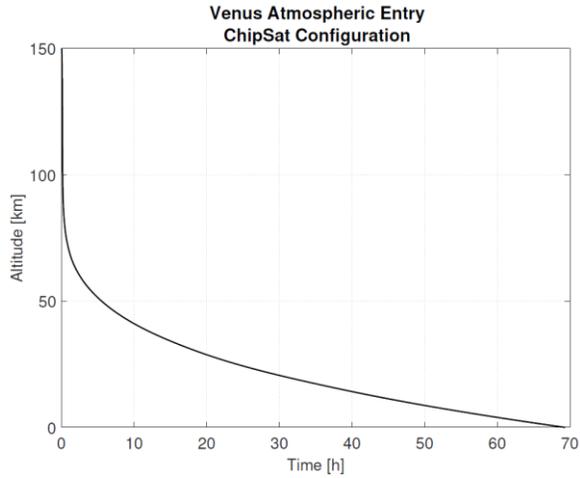

*Figure 9. Altitude time history*

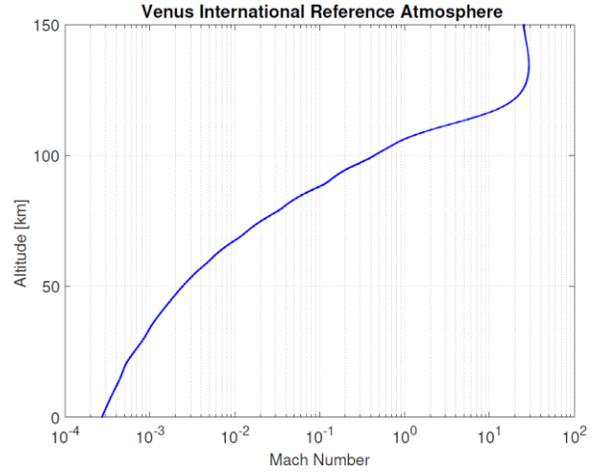

*Figure 12. Mach Number vs Altitude*

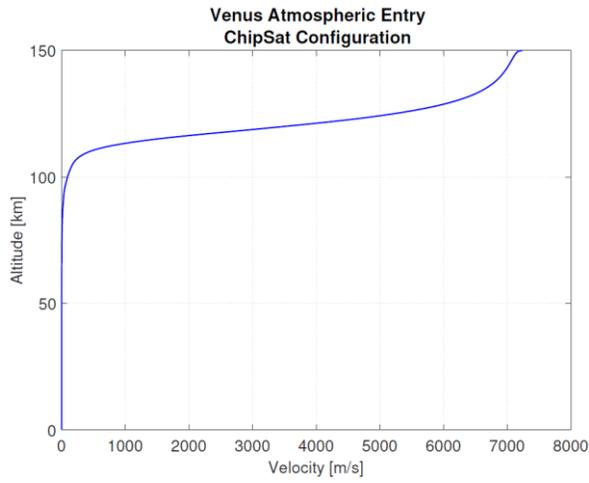

*Figure 10. Velocity vs Altitude*

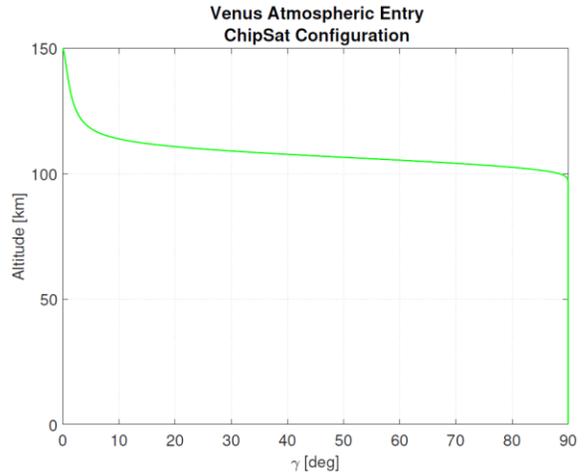

*Figure 13. Flight Path Angle vs Altitude*

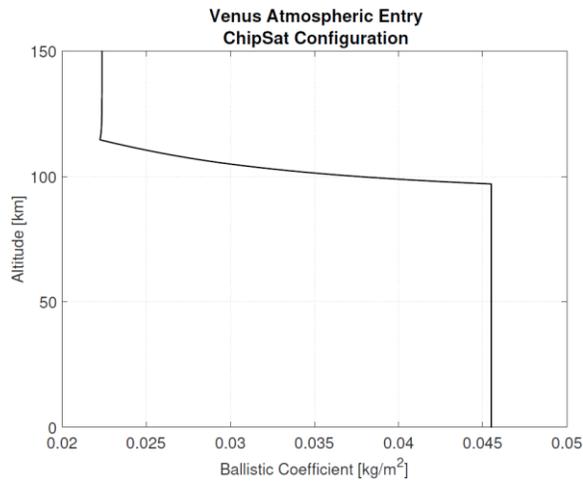

*Figure 11. Ballistic Coefficient vs Altitude*

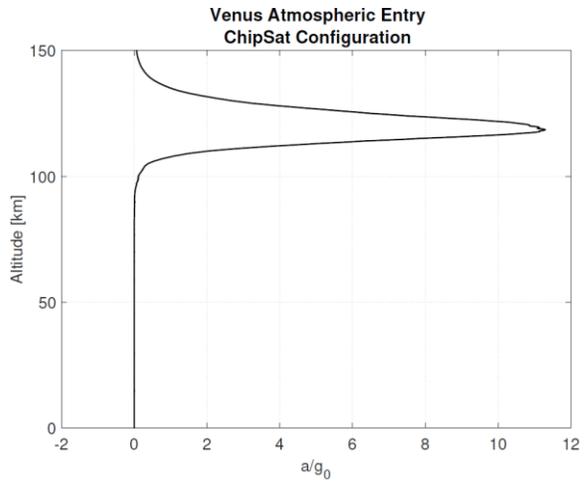

*Figure 14. Acceleration load vs Altitude ($g_0 = 8.87\ m/s^2$)*

The atmospheric model used for this simulation is the Venus International Reference Atmosphere (VIRA), described in Section 2.1. A face-on configuration ($\alpha = 90°$) is considered, therefore the lift force effects are negligible. Concerning the drag coefficient, a $C_D$ profile varying with the flow regime (described in Section 2.2), is used.

The altitude time history in Figure 9 shows that the descent of the ChipSat through Venus' atmosphere is rather long compared to an Earth entry performed with the same spacecraft (see Figure 9). Indeed, the ChipSat reaches Venus' surface after about 69 hours starting the descent from an altitude of 150 km, whilst it reaches Earth's surface in about 6 hours from an altitude of 350 km.

This result is reflected in the velocity trend in Figure 10, showing a dramatic drop of the spacecraft velocity at high altitudes. At an altitude of 100 km, the ChipSat reaches already a quite low velocity value of 94 m/s after only 9.8 minutes. Thereafter, the descent is much slower, and the velocity reduces more gradually, reaching 1 m/s at an altitude of 56 km.

The velocity trend is due to much higher values of the drag coefficient at higher altitudes, implying lower values of the ballistic coefficient in the upper layers of Venus' atmosphere, as shown in Figure 11, which cause a higher velocity gradient $dv/dh$ (see equation 2.3.c).

Figure 12 represents the Mach number profile with respect to the altitude. Defined as $Ma = v/a_s$, where $v$ is the spacecraft velocity and $a_s$ is the speed of sound, the Mach number is an important non-dimensional quantity (together with the Knudsen number), which enables defining the flow regime the spacecraft experiences during the descent. Figure 12 shows that the sonic limit of $Ma = 1$ is reached already at an altitude of 106 km. From Figure 3 in Section 2.1, we deduce that the ChipSat is in a continuum flow regime ($Kn < 0.1$) starting from an altitude of 98 km, at which the flow velocity is already subsonic. These two results are in line with the assumption of considering the constant value $C_D = 1.28$ [17] for the drag coefficient model in continuum, which is valid only for subsonic regimes.

Moreover, Figure 13 represents the flight path angle profile varying during the descent. The final value $\gamma = 90°$ is reached already at an altitude of about 98 km.

Finally, Figure 14 shows the acceleration load profile varying with the altitude, which reaches a peak of 11.3 g at an altitude of about 119 km.

The innovative nature of this study implies that little benchmark results exist in literature. Therefore, the reliability of the results discussed in this section can be justified only from a qualitative perspective, by comparing the trends of the parameters profiles for a Venus atmospheric entry, with the corresponding profiles for an Earth atmospheric entry performed with the same ChipSat, which have been presented in Section 3.2. The main difference is that the entry duration is much longer for a mission on Venus, and this is probably due to the different atmospheric conditions at low altitudes. Indeed, the surface density and pressure values on Venus are respectively about 53 times and 92 times the Earth's surface density and pressure values.

### 4.2 ChipSat Thickness Parametric Analysis and Useful Time Evaluation

A recent study published in the journal *Astrobiology* by Limaye et al. [9] states that the lower cloud layer of the Venusian atmosphere could be an good target for searching for microbial lifeforms, due to the favorable atmospheric conditions (~60 °C temperature, ~1 atm pressure, a not prohibitive UV radiation flux) and the presence of micron-sized sulfuric acid aerosols.

In order to assess the feasibility of a Venus atmospheric entry mission performed with a ChipSat, aiming at providing in-situ scientific studies to search for life in the Venusian atmosphere, a parametric analysis has been performed using the AEM.

Based on this recent astrobiology study and on the potential presence of "unknown ultraviolet absorbers" at the same altitudes already proposed by Grinspoon et al. [11], the range of altitude selected for the in-situ investigation of Venus' atmosphere for microbial lifeforms is **47 - 51 km**. which corresponds to the lower cloud layer.

The parametric analysis investigates the optimal thickness of the ChipSat for the proposed mission objectives. This dimension affects the mission profile strongly, especially the useful time the ChipSat would spend at the selected altitude range (47 – 51 km), to provide data for the scientific studies.

The ChipSat configuration chosen for the simulations is the same used for the analyses presented in Sections 3.2 and 4.1 (see Table 5), except for the ChipSat's thickness, which is the parameter varied during the simulations. A face-on configuration ($\alpha = 90°$) is considered, and the same assumptions made in Section 4.1 for initial values and drag coefficient are taken into account.

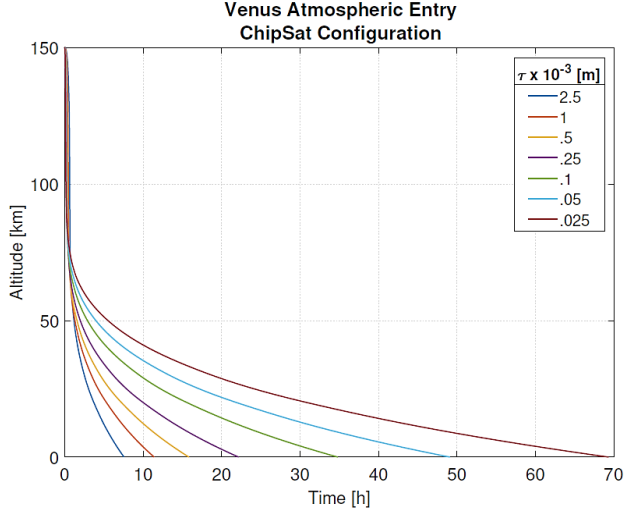

*Figure 15. Altitude time history at different thickness values*

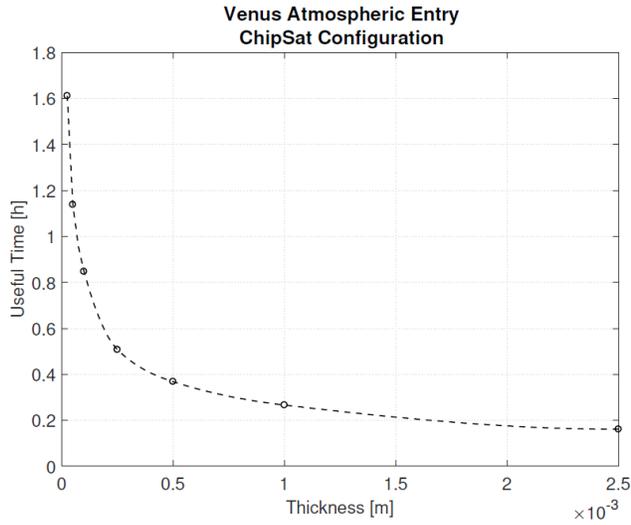

*Figure 16. Useful Time vs spacecraft's thickness*

Figure 15 shows the altitude time history considering different values of the spacecraft's thickness, whilst Figure 16 represents the variation of the useful time varying the spacecraft's thickness.**Error! Reference source not found.** reports the useful time the ChipSat spends at the selected altitude range (47 – 51 km) for different values of the spacecraft's thickness, varying from $2.5\ mm$ to $25\ \mu m$.

*Table 8. Useful Time at 47-51 km for different values of the ChipSat's Thickness*

| CHIPSAT'S THICKNESS [× $10^{-3}$ METERS] | USEFUL TIME [HOURS] |
|---|---|
| 2.500 | 0.16 |
| 1.000 | 0.27 |
| 0.500 | 0.37 |
| 0.250 | 0.51 |
| 0.100 | 0.85 |
| 0.050 | 1.14 |
| 0.025 | 1.61 |

The results of the parametric analysis highlights that a thinner ChipSat would spend a significantly longer time at the selected altitude range. Thus, assuming a 25 µm thickness and a 1 cm side-length as the dimensional limits for the development of such a small spacecraft, the minimum thickness of 25 µm would assure an optimized useful time of about **97 minutes**.

Another important aspect to be evaluated to assess the feasibility of such a mission is the total heat load the ChipSat experiences during entry. The AEM does not include yet a validated aerothermodynamics model, nevertheless Atchison, Manchester and Peck [5] demonstrate that a ChipSat with the same dimensions as the one used in this paper can survive an Earth atmospheric entry. Moreover, Dalle and Spangelo [19] report that a ChipSat with a 32 µm thickness and a 1 cm side-length would survive a descent into the atmosphere of Venus with some minor heat shielding, such as aerogel. Therefore, the baseline configuration that was presented in this paper is expected to follow similar trends, since thinner ChipSats are subjected to lower thermal loads.

Furthermore, the simulation is performed considering a face-on configuration($\alpha = 90°$). A randomly tumbling shaped object would probably experiences a faster descent, due to lower values of the drag coefficient and higher values of the ballistic coefficient in continuum regime. This might motivate the development of an attitude control system or aerodynamic shapes for such a small spacecraft to improve their performance.

Finally, it is important to mention that the AEM does not include a model for the hyper-velocity winds that speed up in Venus' atmosphere faster than the rotation of the planet itself. There is no evidence of how the winds could affect a ChipSat entry, but from a purely theoretical perspective, they could have a beneficial effect for an edge-on configuration (if we do not consider tumbling), making the ChipSat float in Venus' clouds longer than expected, further increasing the useful time for scientific studies. On the other hand, the winds could reduce the entry duration for a face-on configuration.

## 5. Mission Concept and Preliminary Design Considerations

The mission concept proposed to perform in-situ investigation of the lower cloud layer of Venus' atmosphere is described at high level in this section, together with a discussion on the potential design of the spacecraft.

### 5.1 Mission Concept

The proposed idea for the mission concept consists in launching a swarm of ChipSats on-board a mother-spacecraft that would transport them towards Venus, and then deploy them in the planet's atmosphere, similarly to the deployment of the Sprites ChipSats by the Kicksat-2 CubeSat carrier.

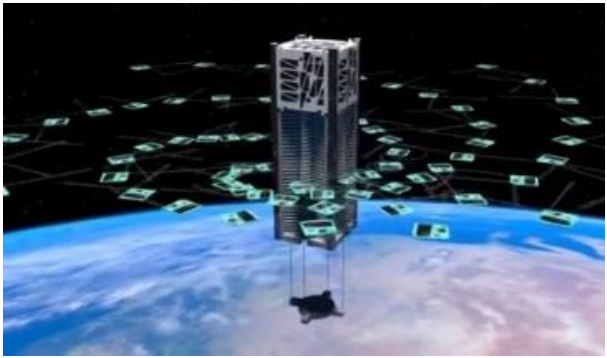

*Figure 17. Artistic impression of the Sprites deployment* [4]

The mother-spacecraft would be injected into an elliptical orbit around Venus with a periapsis between 250 and 150 km, similar to the orbit that two past missions followed for their coasts to Venus, such as the NASA Pioneer Venus Orbiter and the ESA Venus Express. Thus, in proximity of the periapsis the mother-spacecraft would deploy the ChipSats. The small satellites would then descend towards Venus' surface, spending around 97 minutes of useful time at the selected altitudes (47-51 km) to perform their scientific measurements and send back data to the mother-spacecraft, before diving into the lower layers of Venus' atmosphere.

Based on the results discussed in Section 4, the target ChipSat for the mission has the following specifications, listed in Table 9.

*Table 9. ChipSat specifications*

| | |
|---|---|
| Side Length | **1 $cm$** |
| Thickness | 25 $\mu m$ |
| Material Density | 2330 $kg/m^3$ |

### 5.2 Structures

The ChipSat structure would be based on a Printed Circuit Board (PCB), which provide structural integrity and include circuits transmitting signals/current, and Surface Mounted Devices (SMD), components soldered on it. Both these technologies can be downscaled to a 1 cm size ChipSat [20].

### 5.3 Communications

One of the biggest challenge of designing a ChipSat mission is performing communications with these small satellites. Indeed, without the ability to accommodate a high-gain antenna, the data rate from is substantially lower than larger spacecraft with more power availability and directed, high-gain antennas [3].

In the context of the mission concept proposed, the communications subsystem shall allow the ChipSat to send back data to the mother-spacecraft while the scientific measurements at the selected altitude range are performed.

The initial baseline for the communications subsystem would be the 25 mW ISM-band radio and embedded PCB antenna, used on-board the Monarch ChipSats. According to [3], the Shannon Limit for a Monarch in Earth orbit is about 84 kilobits per second. Assuming a similar value for our mission, the potential transmission rate for the ChipSats is quite low. However, it is important to mention that a swarm of hundreds or thousands of ChipSats would be deployed simultaneously. Therefore, "the total data rate is competitive with large, high-power spacecraft" [3]. Moreover, the ChipSats would send data from different locations in Venus' atmosphere, which enables a higher spatial resolution compared to receiving large amounts of data from the same region.

### 5.1 Power

Although Landis and Haag [21] demonstrate that the use of photovoltaics for missions on Venus' lower atmosphere and surface is feasible despite high temperatures, low light levels, and a spectrum that is deficient in the short-wavelength part of the spectrum, the dimensional constraints of a ChipSat require the usage of different power systems, such as supercapacitors.

Compared to Li-ion batteries, supercapacitors are characterized by a higher specific power

($\sim 10^5$ W/kg versus $\sim 10^3$ W/kg), and a lower specific energy ($\sim 10^{-2}$ Wh/kg versus $\sim 10$ Wh/kg) [22]. Thus, they are suitable for our mission, which requires high power for a short period (~97 minutes).

The miniature graphene-based supercapacitors described by Djuric et al. [23] are chosen in this study to supply power to the ChipSat. They are characterized by a length of 6.8 mm, and a 210 µm width of the electrodes, separated from each other by 60 µm. Furthermore, they deliver an area capacitance of 80.5 µF/cm$^2$. These supercapacitors would be fully charged before the deployment, storing the energy to be provided to the ChipSat during its nominal operations.

*5.2 Thermal Protection*

No active thermal protection system is needed to withstand the thermal loads the ChipSat experiences during entry. As per [19], the use of silica aerogel, a low density, light weight material, with high insulating capability [24], is sufficient to protect the spacecraft during the entry.

*5.3 Payload*

To gain insights into the potential existence of lifeforms in Venus' clouds, several different instruments could be used. For in-situ atmospheric measurements, [9] proposes the following devices:

- Compact Raman Lidar to measure organic/inorganic composition, described in [25].
- Fluorescence Lidar to measure organic/biomolecular fluorescence.
- Life detection microscope, described in [26], to detect the presence of microbial cells.

A miniaturized version of these instruments, so that they could be integrated together in a micro-scale spacecraft such as the ChipSat proposed, is not currently available.

An alternative approach is provided by the Miniaturized Multi-Spectral Instrument, proposed by the NASA Langley Research Center team in collaboration with the University of Hawaii for the characterization of planetary surfaces and atmospheres [27]. It is a small-size device (8 – 10 kg, 30 x 30 x 25 cm) that integrates a Raman fluorescence spectrograph and a compact Lidar multi-sensor system, capable of investigation and identification of minerals, organic, and biogenic materials, as well as to perform general atmospheric studies. Although this is a miniaturized payload, the mass and the size of this device still exceed the limits imposed by the ChipSat configuration proposed. Therefore, further technological developments are needed to make it available for such a small spacecraft.

Another alternative is the concept of Lab-on-a-Chip (LOC), a device that integrates several laboratory functions on a single integrated circuit with the maximum size of few square centimeters, able to handle minute sample volumes. The concept proposed by [28] includes the potential development of a Lab-on-a-Chip for Liquid Chromatography - Mass Spectrometry (LOC LC-MS). LC-MS is a technique used to detect metabolites, organic substances naturally occurring from the metabolism of a living organism and that do not directly come from gene expression. It could be very useful to the ChipSat to detect potential biomarkers in Venus' clouds [29].

Finally, the integration of miniaturized versions of cameras (Visible, IR, UV) and magnetometers could lead to unexpected discoveries by providing additional data, such as images of the interior of Venus' atmosphere, as well as measurements that could help to understand better the induced magnetosphere of the planet.

**6. Conclusions and Recommendations**

An *Atmospheric Entry Model* (AEM) has been developed. It is able to compute the planetary entry conditions of a spacecraft of specified geometry in the atmosphere of any planet for which sufficient atmospheric data are available. An overview of the mathematical models implemented in the AEM are described in detail in this paper. The model's results have been verified for spacecraft translational dynamics in the case of an atmospheric entry at Earth, considering both a traditional-sized spacecraft and a ChipSat spacecraft configuration.

Relying on the positive outcome of the results verification, a simulation for a ChipSat entry in Venus' atmosphere has been performed. The reliability of the results is stressed by the similarity between the trends of the parameters profiles for a Venus atmospheric entry, with the corresponding profiles for an Earth atmospheric entry performed with the same ChipSat.

Furthermore, following the selection of the 47 - 51 km altitude range for Venus' atmosphere parametric analysis has been performed using the AEM. The results highlight that a thinner ChipSat would have a significantly longer useful time. Indeed, considering a ChipSat with a 25 µm thickness and a 1 cm side-length, the useful time for the scientific investigation would be 97 minutes. No thermal analysis has been performed, but results in literature suggest that a ChipSat with similar characteristics would survive a descent into the atmosphere of Venus with only minor heat shielding.

Moreover, a mission concept to perform in-situ investigation of the lower cloud layer of Venus' atmosphere has been proposed, and preliminary design considerations regarding the structures, communications, and power have been discussed. Different kinds of payload have been explored for integration on the ChipSat:

- The Miniaturized Multi-Spectral Instrument proposed by [27], which requires further technological developments to meet the size and mass constraints, imposed by the microscopic configuration of the spacecraft.
- The Lab-on-a-Chip for Liquid Chromatography - Mass Spectrometry (LOC LC-MS) proposed by [28].
- Miniaturized cameras and magnetometers for general atmospheric studies.

Finally, some recommendations for future potential developments are discussed below:

a. Consideration of the rotational motion of the ChipSat during entry through an additional ODE.
b. Implementation in the *Atmospheric Entry Model* of a proper aerothermodynamics model to compute the thermal loads the ChipSat experiences during entry.
c. Implementation of a proper model for the high-speed winds of Venus' atmosphere.

**Acknowledgments**

Special thanks to *Initiatives for Interstellar Studies* to award the main author of this paper the *Alpha Centauri Price* for the best Individual Project carried out at the *International Space University*. It provided the right motivation energy to carry on with the project and write this paper for the International Astronautical Congress 2019.